\documentclass[aps,twocolumn,superscriptaddress,showpacs]{revtex4b4}
\usepackage{epsf,latexsym}
\usepackage{times}

\newcommand{\ket}[1]{| #1 \rangle}

\newcommand\N{{\mathrm {I\!N}}} 

\newcommand{\ra}{{\rightarrow}}
\newcommand{\be}{\begin{equation}}
\newcommand{\ee}{\end{equation}}

\def\bbbone{{\mathchoice {\rm 1\mskip-4mu l} {\rm 1\mskip-4mu l} 
{\rm 1\mskip-4.5mu l} {\rm 1\mskip-5mu l}}} 
 
\newcommand{\putfig}[2]{$$\leavevmode\hbox{\epsfxsize=#2 cm 
\epsffile{#1.eps}}$$}

\begin{document}

\title{Quantum Computation with Ballistic Electrons}
\author{Radu Ionicioiu}
\affiliation{Department of Engineering, University of Cambridge, Trumpington Street, Cambridge CB2 1PZ, UK}
\affiliation{Istituto Nazionale per la Fisica della Materia (INFM)}
\affiliation{Institute for Scientific Interchange (ISI), Villa Gualino, 
Viale Settimio Severo 65, I-10133 Torino, Italy}
\author{Gehan Amaratunga}
\author{Florin Udrea}
\affiliation{Department of Engineering, University of Cambridge, Trumpington Street, Cambridge CB2 1PZ, UK}

\begin{abstract}
We describe a solid state implementation of a quantum computer using ballistic single electrons as {\em flying qubits} in 1D nanowires. We show how to implement all the steps required for universal quantum computation: preparation of the initial state, measurement of the final state and a universal set of quantum gates. An important advantage of this model is the fact that we do not need ultrafast optoelectronics for gate operations. We use cold programming (or pre-programming), i.e., the gates are set before launching the electrons; all programming can be done using static electric fields only.
\end{abstract}

\maketitle

In recent years quantum information processing emerged as an important field for theoretical and experimental investigation \cite{ddv_nature}. Using quantum mechanical phenomena for storing and manipulating information it is possible to outperform classical algorithms \cite{shor, grover}. This motivated the present ``gold rush'' for actual physical implementations. There are different proposals for building a quantum computer ({\em quputer}, for short). These include ion traps, NMR quantum computation, cavity QED and single photonics. Among these a solid state implementation of a quputer \cite{kane}-\cite{barnes} has some advantages, including scalability, miniaturization and flexibility in design. A recent experimental result is the control of a qubit using a superconducting Cooper-pair box \cite{nakamura}.

In this article we extend and analyze our model for quantum computation with ballistic electrons proposed in \cite{quputer}. The main idea is to use ballistic electrons as {\em flying qubits} in 1D quantum wires used as electron waveguides. Several requirements have to be met by any implementation of a quputer (DiVincenzo's checklist \cite{ddv3}): {\bf (i)} well defined qubits; {\bf (ii)} low decoherence; {\bf (iii)} initial state preparation; {\bf (iv)} final state measurement; {\bf (v)} universal set of quantum gates. We show how to implement all these steps with ballistic electrons.

\begin{figure}[h]
\putfig{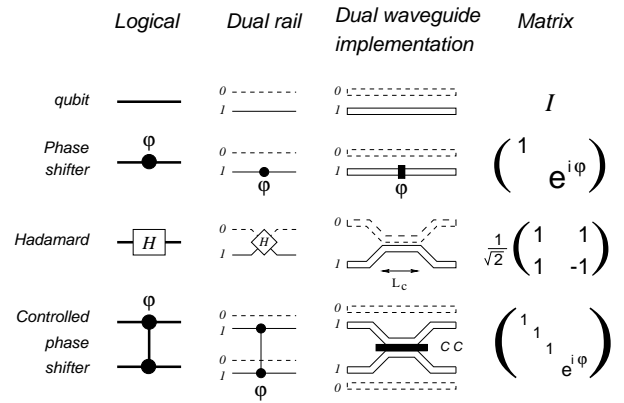}{8}
\caption{Single and dual rail representations for qubits; the {\bf 0}-rails of each qubit are dashed for clarity (the dual rail Hadamard gate cannot be factorized in single rail representations).}
\label{allgates}
\end{figure}

\noindent
{\bf (i) The qubit}\\
Our physical qubit consists of two adjacent 1D quantum wires, called the {\bf 0}- and the {\bf 1}-rail, respectively (dual rail representation \cite{simple_qc}). We define the logical state $\ket{0}$ by the presence of a single electron of energy E$_{\bf k}$ in the {\bf 0}-rail and the logical state $\ket{1}$ by the presence of a single electron (with same energy) in the {\bf 1}-rail. How realistic is this situation? For a semiconductor at low temperatures, the electron density in the conduction band is due to impurities ionization. For a donor concentration of $10^{13}$ cm$^{-3}$, the density of electrons in the conduction band for intrinsic GaAs is $\sim 10^{-5}$ cm$^{-3}$ even at 1 K; therefore a single electron injected in the conduction band is clearly distinguishable from the no electron state. A correspondence between single and dual rail representations is shown in Fig.~\ref{allgates}.

\noindent
{\bf (ii) Coherence}\\
An essential requirement for any implementation of a quputer is to maintain the coherence of the qubits during the entire period of computation. Since we use ballistic single electrons in 1D nanowires, their phase coherence is preserved. The parameter which characterizes the coherence of the system is the {\em phase coherence length} $L_\phi$ over which the electrons maintain their phase coherence. At low temperatures (around 10 mK) the phase coherence length is of the order of tens of microns. For GaAs/AlGaAs heterostructures $L_\phi\sim 30-40\mu$m \cite{gaas}.

Recent experimental \cite{cnt1} and theoretical \cite{cnt2} work demonstrate that metallic single-walled carbon nanotubes (CNTs) can act as long ballistic conductors (quantum wires) over micron lengths even at room temperature \cite{ballistic_CNT}. Simple carbon nanotube devices, like the Y-junction \cite{yjunction, yCNT} and the field effect transistor \cite{cnt_transistor}, have been experimentally demonstrated. Due to rapid advances in the fabrication and manipulation of CNTs, this technology could also be used in the near future to implement the present proposal.

\begin{figure}[h]
\putfig{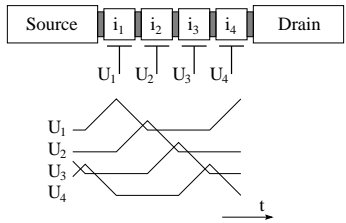}{4.5}
\caption{A single electron pump (SEP).}
\label{sep}
\end{figure}

\noindent
{\bf (iii) Initial state preparation}\\
We prepare the initial state (e.g.~$\ket{0,0,\ldots}$) by injecting a single electron with energy E$_{\bf k}$ in the {\bf 0}-rail of each qubit. We use a single electron pump (SEP) (shown schematically in Figure \ref{sep}), followed by an energy filter.
The SEP works as follows: between a source and a drain there are several conducting islands separated from each other (and from the source and the drain) by tunneling barriers. The size of the islands is typically of the order of tens of nanometers. Due to the Coulomb blockade effect, only one excess electron can be on an island at any time. This can be explained as follows. Due to their size, the islands have a very small capacitance and thus, an elementary charge $e$ on one island induces a large Coulomb potential. Therefore, a second electron is prevented from transferring onto the island before the first electron reaches the neighboring island (or the drain). We now apply a periodic pulse on each gate $U_1,\ldots,U_4$; the pulse on each gate $U_i$ is slightly retarded from the previous one $U_{i-1}$, such that these form a traveling wave which pushes the electron from the source to the drain. A typical frequency for the drive signal on the gates is between 10-100 MHz \cite{sep}.

Another essential role of the SEP is to synchronize different qubits (i.e.~different branches of the calculation). By adjusting the timing between the gate pulses $U_1,\ldots,U_4$ we can make two electrons of different qubits to arrive simultaneously at the interaction region (the 2-qubit gate described below). Since the electrons injected in different qubits should be synchronized at all times during the computation, they need to have the same wave vector $\bf k$ (and hence the same energy E$_{\bf k}$). This is done using a double potential barrier as an energy filter. The electron can tunnel through the double barrier if and only if its energy is equal to the energy of the bound state inside the barrier (resonant tunneling effect). By adjusting the height of the potential inside the double barrier, we can have in principle a tunable energy filter for electrons.

\noindent
{\bf (iv) Final state measurement}\\
At the end of the calculation we need to measure the state of each qubit. Each qubit rail is coupled to a single electron transistor (SET), which is sensitive to single electron charges \cite{set, measure-set}. A SET also uses the Coulomb blockade effect. If the source-drain voltage is just above the Coulomb threshold, the source-drain current is very sensitive to the gate voltage. By having the qubit rail as the gate of the SET, the source-drain current will be modified by the presence of the electron wave. One problem with this scheme is that we need to trap the electron at the end of the rail, since a SET needs a relatively long time $\sim 10^{-7}$s to measure the charge. A possible solution is to use a turnstile at the end of each rail, just before measurement. Very recently, a coherent single-electron turnstile operating in a picosecond time scale has been proposed \cite{turnstile} (the coherence is essential here).

\noindent
{\bf (v) Quantum Gates}\\
Any quantum computer can be build using only single- and two-qubit gates \cite{gates}. We choose the following universal set of quantum gates: $\left\{ H, P_\varphi, CP_\pi \right\}$, where $H=\frac{1}{\sqrt{2}}\pmatrix{1&1 \cr 1&-1}$ is a Hadamard gate, $P_\varphi = \mbox{diag}\,(1,\, e^{i\varphi})$ is a single-qubit phase shift, and $CP_\pi$ is a controlled sign flip. We shall use the more general two-qubit gate $CP_\varphi = \mbox{diag}\, (1,\,1,\,1,\,e^{i\varphi})$. Numerical simulations of these quantum gates have been presented by Bertoni {\it et al.}~\cite{bertoni}.

\begin{figure}
\putfig{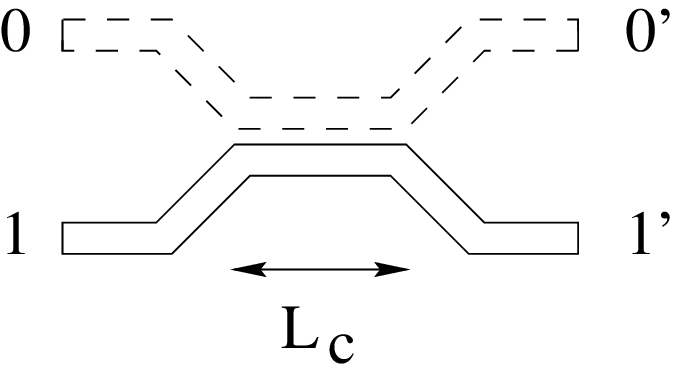}{2.5}
\caption{Electron beam splitter based on an electron waveguide coupler.}
\label{coupler}
\end{figure}

\noindent
{\bf Hadamard gate} -- A Hadamard gate is equivalent to a beam splitter. This is implemented with an {\em electron waveguide coupler} \cite{alamo}-\cite{nl_coupler} which is formed from two parallel waveguides brought together to an interaction region of length $L_c$, as in Figure \ref{coupler}. Since we work in a dual rail representation, only one electron at a time will be either in the {\bf 0}- or in the {\bf 1}-rail. As the electron propagates along the waveguide, it oscillates back and forth between the two (due to the evanescent coupling between the waveguides). After a transfer length $L_t$ (equivalent to the half period of the oscillation), the electron injected initially in the upper waveguide is totally transferred in the lower one. In order to use this device as a symmetric beam splitter, we chose the coupling length of the gate to be half of the transfer length $L_c=L_t/2$.
In reference \cite{tsukada}, the transfer length for the complete electron transfer from one waveguide to the other is found to be $0.28\mu$m (for a separation of the two wave guides of 0.2$\mu$m and a coupling energy $\Delta E=10$ meV); therefore, a beam splitter can be made as small as $0.14\mu$m. This is also within the limits of currently available optical lithography methods for pattern definition.

\begin{figure}
\putfig{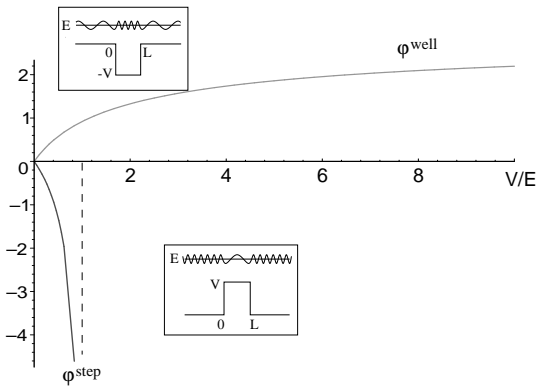}{8}
\caption{The phase shift of an electron wave incident on a quantum well (top) and on a step potential (bottom) for $n=1$. The insets show schematically the two cases.}
\label{bigfig}
\end{figure}

\noindent
{\bf Phase shift} -- this is implemented using a potential barrier with height smaller than the electron energy $V< E (=\mbox{E}_{\bf k})$. In order to have no reflection from the potential step, the width $L$ of the barrier should be a multiple of the half wavelength of the electron in the step region, $L={n\over 2}\lambda$, $n \in \N$. A simple calculation shows that in this case the emergent wave function of the electron has a phase shift relative to the incident one, $\psi_{out}= e^{i\varphi}\ \psi_{in}$, with the phase shift given by
\[ \varphi^{step}= n\pi \left( 1- \frac{1}{\sqrt{1-V/E}} \right) \ \ \ , \ \ \ n \in \N \]
The same effect can be achieved with a quantum well instead of a quantum step, but the sign of the phase shift is now reversed (see Fig.\ref{bigfig}).

\[ \varphi^{well}= n\pi \left( 1- \frac{1}{\sqrt{1+V/E}} \right) \ \ \ , \ \ \ n \in \N \]
As we can see from Figure \ref{bigfig}, for the step potential there is a vertical asymptote for $V=E$ and the induced phase shift tends to infinity. For the quantum well, however, there is a horizontal asymptote at $\varphi=n \pi$. In this case the phase shift is smaller, but more stable under variations of $V$ than in the step potential case, where a small perturbation in the potential can induce a large variation in the phase shift. The previous argument assumed that the electron has a well defined wavelength and energy (plane wave). In reality the electron is described by a quasi-monoenergetic wave-packet with a wavelength distribution sharply peaked around $\lambda=h/\sqrt{2m^*E}$. In this case there will be a small reflected component, but this can be made sufficiently small by an appropriate design. The same problem appears in quantum optics. Although a single photon is represented by a wave packet with a sharply defined wavelength, there are anti-reflective coatings with high efficiency so that the reflected component can be neglected.

\noindent
{\bf Controlled phase-shift} -- we use a Coulomb coupler (CC) described by the Hamiltonian \cite{c_coupler}:
\be
H= \hbar \chi N_A N_B
\ee
where $\chi$ is the coupling constant and $N_A$, $N_B$ are the particle number operators for the two qubits, $N_A= a^\dagger a$, $N_B= b^\dagger b$. If $t$ is the interaction time, the effect of the gate on the two fields is:
\[ a\ \ra\ a'=a \ e^{-i\chi t N_B}  \ \ \ ,\ \ b\ \ra\ b'=b \ e^{-i\chi t N_A} \]
and thus the two electrons give each other a mutual phase modulation proportional to the particle number in each field. Since in our case $N_i=0$ or 1, the action of the gate can be written as
\[ \ket{00}\ra\ket{00} \ \ \ ,\ \ \ket{01}\ra \ket{01} \]
\be
\ket{10}\ra\ket{10} \ \ \ ,\ \ \ket{11}\ra e^{-2i\chi t}\ket{11}
\ee

The phase induced on each electron $\varphi=-\chi t$ is proportional to the coupling constant and to the interaction time, and hence to the gate length. By making the coupling $\chi$ sufficiently strong, the length of the gate can be made in principle less than $1\mu$m.

In order to have a strong enough coupling between the two electrons in the Coulomb coupler, the two qubit rails should be sufficiently close. On the other hand, the tunneling probability between the two qubits should be negligible. Therefore, a Coulomb coupler should have a high potential barrier in order to prevent electrons from tunneling from one qubit rail to the other, and in the same time the two rails should be close enough in order to make them interact. The tunneling amplitude is suppressed by a factor of $\exp(-L\sqrt{2m^*(V-E)}/\hbar)$, where $L$ and $V$ are the width and height of the barrier, respectively. Thus, we can keep the tunneling amplitude constant and small by making $V$ large enough in order to prevent the tunneling, while making $L$ sufficiently small in order make the electrons interact.

Another way of making the tunneling probability small enough while keeping the two electron rails close is to use the resonant tunneling effect. Thus, if we have a double potential barrier between the two electron waveguides, the tunneling is inhibited at certain energies of the incident wave. An electron can tunnel only if its energy is the same as one of the bound states of the well. By making these two energies very different, we can inhibit the tunneling between the two waveguides.

\begin{figure}
\putfig{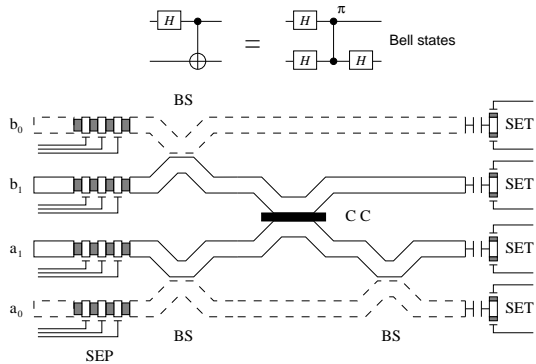}{7}
\caption{Producing entanglement: a quantum network for the Bell states. The beam splitters (BS) are electron waveguide couplers and the Coulomb coupler (CC) is used to entangle qubit $a$ and $b$; the {\bf 0}-rails of each qubit (a$_0$ and b$_0$) are dashed for clarity (the energy filter on each qubit rail is ommited for simplicity).}
\label{bell}
\end{figure}

Since these gates are universal, any quantum algorithm can be built using only these three types of gates. A two-qubit gate between arbitrary qubits is executed by swapping qubits until they become neighbors, perform the gate, and swapping them back to their initial positions (a {\sf SWAP} gate is constructed out of three {\sf CNOT}s). A quantum network for producing entangled (Bell) states is presented in Fig.~\ref{bell} (see also \cite{ri_bell}).

We discuss now the advantages of the proposed model. Both the logical $\ket{0}$ and $\ket{1}$ state have the same energy, and therefore their time evolution is identical. Moreover, they are both stable (no spontaneous decay), since they correspond to the double degenerate ground state (electron in the {\bf 0}- or in the {\bf 1}-rail). As the two rails are well separated, there is no tunneling between them outside the gate regions. This situation is different from proposals where the $\ket{1}$ state is an excited state of the system. In this case two problems occur: (i) the $\ket{1}$ state is not stable (it has a finite life-time due to spontaneous decay); (ii) due to the time evolution, $\ket{1}$ picks up a phase $e^{-it\Delta E/\hbar}$, with $\Delta E$ the energy difference between the excited and the ground state.

Another important advantage of our model is the fact that we do not need ultrafast electronics (or laser pulses) for gate operations. Due to the short decoherence times, other solid state proposals use ultrafast electronics or laser pulses (on the femtosecond scale) in order to perform gate operations subdecoherently. In our proposal we use {\em cold programing} (or pre-programming): all the gates are set in advance before ``launching'' the electrons. Programming is done by switching on/off the gates situated along the quantum wire according to the algorithm to be executed. The phase shift gate $P_\varphi$ is the easiest example: it can be turned on and off by simply turning on/off the potential applied to the gate situated on top of the quantum wire. One way of turning off the Hadamard gate is by making the potential barrier between the two rails high enough such that the tunneling between the rails is inhibited. The other way is to use the resonant tunneling beam-splitter proposed in \cite{resonant_bs}: the electron can tunnel only if its energy is equal to that of the bound state of the double barrier. By varying an external potential the beam-splitter can be thus turned on and off. Finally, the controlled phase shifter gate $CP_\varphi$ can be designed similarly; for $\varphi=2\pi$, the gate is equivalent to the unity $CP_{2\pi}= \bbbone$. All the gates can be turned on and off by static electric fields and thus the quputer is programmable.

There are some advantages and disadvantages of using single electrons compared to single photons. The main difference between the two is the coupling strength. Due to a very weak coupling, photons have a longer coherence time than electrons, but on the other hand it is much more difficult to make them interact, i.e., to construct a 2-qubit gate which operates at single photon level (we need huge third-order susceptibilities $\chi^{(3)}$). On the other hand it is easier to construct a 2-qubit gate and to prepare and detect single particle states using electrons than photons.

In conclusion, we have shown how to implement all the building blocks required for a solid state quputer using ballistic electrons as {\em flying qubits} in 1D nanowires. We initialize the computer by injecting single electrons of energy E$_{\bf k}$ in one of the rails of each qubit; this is done with a single electron pump (SEP) and an energy filter. Measurement of the final state is performed with a single electron transistor (SET) coupled to the output of each qubit. We implement three types of quantum gates which are universal for quantum computation: a Hadamard gate, a phase shifter and a controlled phase shifter. All these basic elements can be implemented using presently available technology.

\noindent
{\bf Acknowledgments.} We are grateful to Ehoud Pazy, Paolo Zanardi, Fausto Rossi and Irene D'Amico for comments and enlightening discussions.

\end{document}